\documentclass[pra,twocolumn,10pt,amsmath,amssymb,nofootinbib,bm,bbm,notitlepage,superscriptaddress]{revtex4-2}

\usepackage[english]{babel}

\setcitestyle{sort&compress,numbers}
\usepackage{amsthm, color, mathtools, amsmath, amssymb, setspace,bbm, tensor, subfigure, mathtools, placeins, graphicx, wrapfig,float, verbatim, xcolor,enumitem}
\usepackage{graphicx}
\usepackage[unicode=true, breaklinks=false, pdfborder={0 0 1}, backref=false, colorlinks=true, linkcolor=blue, citecolor=blue]{hyperref}
\usepackage{cancel,bm}
\usepackage{bbm}

\let\oldsqrt\sqrt
\def\sqrt{\mathpalette\DHLhksqrt}
\def\DHLhksqrt#1#2{%
\setbox0=\hbox{$#1\oldsqrt{#2\,}$}\dimen0=\ht0
\advance\dimen0-0.2\ht0
\setbox2=\hbox{\vrule height\ht0 depth -\dimen0}%
{\box0\lower0.4pt\box2}}

\newcommand{\tr}[1]{\operatorname{tr}\left(#1\right)}

\newcommand{\tra}{\operatorname{tr}}

\newcommand{\id}{\mathbbm{1}}

\newcommand{\add}[1]{#1}%{{\color{blue}#1}}

\def\bra#1{\mathinner{\langle{#1}|}}

\def\ket#1{\mathinner{|{#1}\rangle}}

\newcommand*\xbar[1]{%
   \hbox{%
     \vbox{%
       \hrule height 0.5pt % The actual bar
       \kern0.5ex%         % Distance between bar and symbol
       \hbox{%
         \kern-0.2em%      % Shortening on the left side
         \ensuremath{#1}%
         \kern-0.0em%      % Shortening on the right side
       }%
     }%
   }%
} 

\def\BraVert{\egroup\,\mid\,\bgroup}

\def\ketbra#1#2{\ket{#1\vphantom{#2}}\!\bra{#2\vphantom{#1}}}

\def\bra#1{\mathinner{\langle{#1}|}}

\def\ket#1{\mathinner{|{#1}\rangle}}

\usepackage{bbm, braket}
\usepackage[mathscr]{euscript}
\allowdisplaybreaks
\newtheorem*{theorem*}{Theorem}

\newtheorem{lemma}{Lemma}

\newtheorem{corollary}{Corollary}
\newtheorem*{corollary*}{Corollary}

\newtheorem*{observation*}{Observation}
\newcommand{\nn}{\nonumber}
\newcommand{\cv}[1]{(#1)}%{\left(#1\right)}
\newcommand{\cvb}[1]{[#1]}%{\left[#1\right]}
\newcommand{\cvc}[1]{\{#1\}}%{\left\{#1\right\}}
\newcommand{\cvv}[1]{\vert #1\vert}%{\left\vert #1\right\vert}

\newcommand{\cvr}[1]{\left\langle #1\right\rangle}
\newcommand{\iden}{\id}

\newcommand{\oper}[1]{\hat{#1}}

\newcommand{\cket}[1]{\left\vert #1\right)}
\newcommand{\cbra}[1]{\left( #1\right\vert}
\newcommand{\cketbra}[2]{\cket{#1}\cbra{#2}}
\newcommand{\cbraket}[2]{\left( #1\vert #2\right)}

\usepackage{orcidlink}

\begin{document}

%\title{When does an entanglement breaking channel break entanglement?}
\title{Bounds on the breaking time for entanglement-breaking channels}
\author{Fattah Sakuldee\,\orcidlink{0000-0001-8756-7904}}
\email{fattah.sakuldee@ug.edu.pl}
\affiliation{The International Centre for Theory of Quantum Technologies, University of Gda\'nsk, Jana Ba\.zy\'nskiego 1A, 80-309 Gda\'nsk, Poland}
\author{{\L}ukasz Rudnicki\,\orcidlink{0000-0001-8563-6101}}
\email{lukasz.rudnicki@ug.edu.pl}
\affiliation{The International Centre for Theory of Quantum Technologies, University of Gda\'nsk, Jana Ba\.zy\'nskiego 1A, 80-309 Gda\'nsk, Poland}
\affiliation{Center for Theoretical Physics, Polish Academy of Sciences, 02-668 Warszawa, Poland}

\begin{abstract}
Entanglement-breaking channels are quantum channels transforming entangled states to separable states. Despite a detailed discussion of their operational structure, to be found in the literature, studies on dynamical characteristics of this type of maps are yet limited. We consider one of the basic questions: for Lindblad-type dynamics, when does a given channel break entanglement? We discuss the finite-dimensional case where the quantification of entanglement via entanglement witnesses is utilized. For the general setup, we use the method of quantum speed limit to derive lower bounds on entanglement breaking times in terms of an input state, the dynamical map and the witness operator. Then, with a particular choice  of the input state and the entanglement witness, the bounds for the breaking time are turned to solely reflect the characteristics of the dynamics. 
\end{abstract}

\date{\today}
		
\maketitle

\section{Introduction}\label{sec:intro}
Quantum speed limit \cite{Mandelstam45,Mandelstam1991,Fleming,Margolus98, Aharonov,Zych} is a well-known fundamental concept related to the question about time-energy uncertainty relations in quantum mechanics. However, at the same time, several diverse applications of quantum speed limit (QSL) can be found  \cite{PhysRevLett.110.050402,PhysRevLett.110.050403,PhysRevLett.120.070401,PhysRevLett.120.070402,PhysRevLett.120.060409,Campaioli2019tightrobust,PhysRevA.103.022210,PhysRevResearch.2.023125,PhysRevLett.126.180603}. The most popular form of QSL is due to Mandelstam and Tamm \cite{Mandelstam45,Mandelstam1991} and depends on the variance of the generator of time evolution. However, sometimes the variance would give an unreasonable assessment, and the average value of the generator is employed, leading to the so-called Margolus-Levitin QSL \cite{Margolus98}, which in fact has been derived before by Fleming \cite{Fleming}. 

One can ask a philosophical question, namely: Why does QSL provide useful pieces of information, given that it is ``just'' a mere consequence of an underlying time evolution? Here we consider a problem which, on top of being interesting and relevant by itself, also perfectly illustrates the way in which QSL overcomes an overall complexity of the full description of a quantum system's dynamics.

To be more precise, we are concerned with quantum channels generated by Markovian dynamics of open quantum systems. We ask whether such channels are entanglement breaking. Any channel $\Phi$ is called entanglement breaking (EB) if the composite state $\cv{\Phi\otimes\mathcal{I}}\cvb{\rho}$ is always separable, even for entangled input states $\rho$ \cite{Horodecki2003}. Since we only consider quantum channels which belong to a semigroup parametrized by $t\geqslant 0$, i.e., channels of the form $\Phi_t=e^{t\mathcal{L}}$, the question we pose gains a bit of structure. In particular, for $t=0$ we have  $\Phi_0=\mathcal{I}$, so the channel $\Phi_0$  \textit{is not} entanglement breaking. We call such channels entanglement preserving (EP). One can expect that, depending on $\mathcal{L}$, the range of time splits into two nonempty and disjoint sets,
\begin{equation}
    t\in[0,\infty[=T_{\mathrm{EP}}(\mathcal{L})\cup T_{\mathrm{EB}}(\mathcal{L}),
\end{equation}
with selfexplanatory interpretations. Even though $t=0\in T_{\mathrm{EP}}(\mathcal{L})$, for every $\mathcal{L}$, both sets  in general will not be connected. It might happen that  the channel is EP for $t\in[0,t_{\mathrm{EB}}[$ and starts to be EB at $t=t_{\mathrm{EB}}$. However, at a later time it is again EP (revival of the EP property).

We might ask whether both sets can be characterized in a more direct way. It is clear that, even for qubit channels, the above task seems hopeless. In this simplest case, even though one is able to explicitly write down cumbersome conditions which describe the EB property \cite{Ruskai2003}, it is not possible to turn such complex inequalities into an informative description of $T_{\mathrm{EB}}(\mathcal{L})$.

In this paper we establish quantum speed limit for (potentially) entanglement-breaking channels, i.e., given the fact that $\Phi_0$ is entanglement preserving, we  bound from below the time in which this property might be lost. In other words, we are to bound the time $t_{\mathrm{EB}}$, such that a given channel $\Phi_t$ is certainly entanglement preserving for $t\in[0,t_{\mathrm{EB}}[$. The standard formulation of QSL tells us what is the minimum time necessary to pass from a given state to an orthogonal state. Our results tell what is the minimal time during which the channel in question is entanglement preserving. We shall stress that this methodology is not aiming at delivering the very exact moment in which the channel becomes entanglement breaking since the problem of separability is, in general \cite{Gharibian}, NP-hard (however, in certain circumstances it can almost always be solved \cite{Hies22}).

The paper is organized as follows. In Sec. \ref{sec:prelim} we present the methodology as well as a necessary formal background concerning quantum channels. We then utilize entanglement witnesses to establish in Secs. \ref{sec:Breaking-time} and \ref{sec:W-Psi} various bounds for $t_{EB}$, in particular, bounds inspired by the Mandelstam-Tamm QSL and Margolus-Levitin QSL.  In Sec. \ref{sec:qubit-eg} we discuss the results with an example.

\section{Preliminaries}\label{sec:prelim}
In accordance with formalism of quantum mechanics, let $\mathcal{H}_1$ be a Hilbert space of a system under consideration and let $\mathcal{H}_2$ be a Hilbert space of an auxiliary system. To study open system evolution, being a $t-$parametrized completely positive and trace-preserving map on $\mathcal{B}\cv{\mathcal{H}_1},$ one can rely on expectation values \cite{sakurai2017,Pollock2018}
	\begin{equation}
		\cvr{A\cv{t}}_\rho = \tr{\oper{A}~\!\cv{\Phi_t\otimes\mathcal{I}}\cvb{\rho}}.\label{eq:OQD-basic}
	\end{equation}
Here $\rho$ denotes an initial state acting on a composite Hilbert space $\mathcal{H}_1\otimes\mathcal{H}_2$, while $\oper{A}$ is an arbitrary observable therein, i.e., a Hermitian operator from $\mathcal{B}\cv{\mathcal{H}_1\otimes\mathcal{H}_2}$. 

Different ways to characterize the map $\Phi_t$ concern different choices of initial states and observables. For instance, in quantum process tomography \cite{Gladden1997,heinosaari2011}, the overall profile of the map $\Phi_t$ is recovered by collecting several $\cvr{A\cv{t}}$ constructed from various pairs of $\rho$ and $\oper{A}$, so that both inputs and observables exhaust some fixed bases of the matrix space $\mathcal{B}\cv{\mathcal{H}_1}$. In this case full knowledge about the auxiliary space $\mathcal{H}_2$ is not required. Another example is the extraction of a characteristic of an external system, encoded in an interaction with the probe system $\mathcal{H}_1$ (signal), where in this case the auxiliary system $\mathcal{H}_2$ can be taken as a reference (idler), a catalyst, or the source of enhancement for the manipulation \cite{Jonathan1999}. 

%define witness
For the sake of studying the dynamics of entanglement,  concerning the action of the map $\Phi_t$,  an entanglement witness  $\oper{W}$ \cite{Terhal2000,Terhal2001,Horodecki1996,Ghne2009} will be an appropriate choice for the observable $\hat A$. With this choice
\begin{equation}
	w_\rho\cv{t} := \cvr{W\cv{t}}_\rho = \tr{ \oper{W}~\!\cv{\Phi_t\otimes\mathcal{I}}\cvb{\rho}} \label{eq:w_t-gen},
\end{equation}
and
\begin{equation}
	w_\rho\cv{0} < 0 \label{eq:init-w-gen},
\end{equation}
since by definition $\tra(\sigma \oper{W})\geqslant 0$ if  $\sigma$ is a separable state, and we require $\rho$ to be entangled.

In a more formal fashion we now recall that $\Phi_t$ is called \emph{entanglement breaking} \cite{Ruskai2003,Horodecki2003} if $\cv{\Phi_t\otimes\mathcal{I}}\cvb{\sigma}$ is separable for all density matrices $\sigma$ acting on $\mathcal{H}_1\otimes\mathcal{H}_2$, where $\mathcal{H}_2$ is a finite dimensional Hilbert space. 
By setting $\rho$ to be entangled, as exposed in Eq.~\eqref{eq:init-w-gen}, and if $\Phi_t$ becomes entanglement breaking at a time moment $t_{\mathrm{EB}}$, this property will be reflected in the sign of the function $w_{\rho}\cv{t_\mathrm{EB}}$. This sign shall change before $t_{\mathrm{EB}}$, or, in an optimal case, $w_{\rho}\cv{t_\mathrm{EB}}=0$.

%connection to CJI
Interestingly, when both subsystems are finite dimensional and symmetric, i.e., $\mathcal{H}_1=\mathcal{H}_2\simeq \mathbb{C}^d$,  
the initial state $\rho$ can be chosen to be a maximally entangled, pure state $\rho=\ketbra{\Psi_+}{\Psi_+}$ \cite{heinosaari2011}, where 
\begin{equation}\label{MaxEntS}
\ket{\Psi_+}=d^{-1/2}\sum_{k}\ket{k}\otimes\ket{k},
\end{equation}
and $\cvc{\ket{k}}$ is a basis of $\mathbb{C}^d$. In such a case, the final state $\rho_{\Phi_t}=\cv{\Phi_t\otimes\mathcal{I}}\cvb{\rho}$ is simply a Choi-Jamio{\l}kowski isomorphic representation of the map $\Phi_t$ \cite{Jamiokowski1972,Choi1975}. In this sense one can say that entanglement of $\rho_{\Phi_t}$ certifies that the corresponding map $\Phi_t$ is entanglement preserving.

For nonsymmetric finite dimensional cases, and also for infinite dimensional case, even though the notion of isomorphism may not be applicable, the basic concepts can be illustrated in the similar way, i.e., an entanglement witness for the outcome state can also be an EP witness for the dynamical map. However, here we shall restrict our attention only to the finite dimension and symmetric situation (both subsystems have the same dimension).

From the perspective of our main question, we are particularly interested in a \emph{configuration-breaking} time $t_{CB}{\cv{\rho,\oper{W}}}$ defined as
	\begin{equation}
		t_{CB}\cv{\rho,\oper{W}} = \min\arg \cvc{w_\rho\cv{t}\geqslant 0} \label{eq:t_B-finding}.
	\end{equation}
In other words, this is minimal time $t$ in which, for a given prepare-measure configuration $\cv{\rho,\oper{W}}$, the state $\rho_{\Phi_t}$ is no longer classified as entangled. Clearly
\begin{equation}
t_\mathrm{EB}=\max_{\cv{\rho,\oper{W}}} t_{CB}\cv{\rho,\oper{W}},
\end{equation}
because on the one hand, for a given $\rho$, we shall find an optimal entanglement witness which works for the longest possible time, while in the second step we need to find the optimal entangled state $\rho$.

The aim of this paper is to find lower bounds for $t_{CB}\cv{\rho,\oper{W}}$, and consequently, bounds for $t_\mathrm{EB}$. From now on, we shall omit the arguments of $t_{CB}$. \add{We note in passing that the time $t_\mathrm{EB}$ is a notion which relies on a possibility of entanglement detection (through a witness or an entanglement measure in general). Therefore, this time cannot by itself play the role of a witness.} 

\subsection{Partitioning by Spectral Basis}\label{sec:spectral}
%generic finite dim; CJI a bit
\begin{table*} 
    \caption{\label{tab:classification}Classification of contributions to the witness function with respect to the characteristics of eigenvalues of the dynamical generator $\mathcal{L}.$ Class I corresponds to the trivial eigenvalue; class II represents pure decay; class III comprises the rest. The index $j$ runs from $1$ to $L=(d^2-K-1)/2$. Note that we express arguments of complex numbers with respect to the branch $\left(-\pi,\pi\right]$.}
    \begin{ruledtabular}
    \begin{tabular}{ccccc}
		Class & $\Gamma_\alpha=\Re\lambda_\alpha$ & $\omega_\alpha=	\Im\lambda_\alpha$ & $r_\alpha$ & $\phi_\alpha$ \\ 
				\hline
		I & $\Gamma_0=0$ & $\omega_0=0$ & $r_0\geqslant 0$ & $\phi_0=\in\{0,\pi\}$ \\ 
		II & $\Gamma_1,\ldots,\Gamma_K\leqslant 0$ & $\omega_1=\ldots=\omega_K=0$ & $r_\alpha\geqslant 0$ & $\phi_1=\ldots=\phi_K\in\{0,\pi\}$ \\ 
		III & $\tilde{\Gamma}_{j}=\tilde{\Gamma}_{2j}\leqslant 0$ & $\tilde{\omega}_{j}=-\tilde{\omega}_{2j}\geqslant 0$ & $\tilde{r}_{j}=\tilde{r}_{2j}\geqslant 0$ & $\tilde{\phi}_{j}=-\tilde{\phi}_{2j}> 0$ \\
	\end{tabular}
    \end{ruledtabular}
\end{table*}

Throughout this article we assume that $\Phi_t$ is rendered by a Lindblad generator \cite{Lindblad1976}, where the map converges to an identity map as the time $t\downarrow 0$ and supposedly becomes entanglement breaking later on. 
In particular, for a given density matrix $\sigma\in\mathcal{B}\cv{\mathcal{H}_1},$ it is a continuous completely positive and trace-preserving map $\Phi_t=e^{t\mathcal{L}}$ on $\mathcal{B}\cv{\mathcal{H}_1}$ with
	\begin{equation}
		\dfrac{d\sigma}{dt} = \mathcal{L}\cvb{\sigma}\label{eq:Lindblad},
	\end{equation}
satisfying $\Phi_{t+s}=\Phi_t\circ\Phi_s$ for $t\geqslant 0$ and $\lim_{t\downarrow 0}\Phi_t=\mathcal{I}$ where $\mathcal{I}$ is an identity map on  $\mathcal{B}\cv{\mathcal{H}_1}.$ 

For further convenience we describe the situation in the Heisenberg picture, where the dual map $\Phi^*_t=e^{t\mathcal{L}^*}$ is unital. 
The generator $\mathcal{L}^*$ can be written as
\begin{equation}
    \mathcal{L}^*=\sum_{\alpha}\lambda_\alpha u_\alpha\tr{ v^\dagger_\alpha\cdot},
\end{equation}
where both $u_\alpha$ and $v_\alpha$ are mutually orthogonal operators satisfying 
\begin{equation}
   \tr{ v^\dagger_\alpha u_{\alpha'}}=\delta_{\alpha\alpha'}.
\end{equation}
This is simply the spectral decomposition on the operational level, where $\mathcal{L}^*\cvb{ u_\alpha}=\lambda_\alpha u_\alpha$ defines the corresponding eigenbasis. 
Hence by spectral theorem 
\begin{equation}
    \Phi^*_t=e^{t\mathcal{L}^*}=\sum_{\alpha}e^{\lambda_\alpha t} u_\alpha\tr{ v^\dagger_\alpha\cdot}. \label{eq:Phi-decom}
\end{equation}  

Since we discuss the case when both parties in the composite system are finite dimensional, the operators can be vectorized and the dynamical map becomes a linear transformation \cite{heinosaari2011}. Let us therefore write $ u_\alpha=\cket{\alpha}$ and $\tr{ v^\dagger_\alpha\cdot}=\cbra{\alpha}$, so that $\mathcal{L}^*=\sum_{\alpha}\lambda_\alpha\cketbra{\alpha}{\alpha}$ and $\Phi^*_t=\sum_{\alpha}e^{\lambda_\alpha t}\cketbra{\alpha}{\alpha}$. We can see that the conditions $\mathcal{L}^*\cket{\alpha}=\lambda_\alpha\cket{\alpha}$ and $\cbraket{\alpha}{\alpha'}=\delta_{\alpha\alpha'}$ define the basis and its dual for the subspace of operators with respect to $\mathcal{L}^*$. 

We stress that, within the introduced vectorization,  $\cket{\alpha}$ does not need to be a Hermitian operator. We also assume that $\cvc{\cket{\alpha}}$ form a basis of $\mathcal{B}\cv{\mathcal{H}_1}$ inducing a resolution\footnote{If $\{\cket{\alpha}\}_\alpha$ does not form the basis, one can restrict the description to the subspace of $\mathcal{B}\cv{\mathcal{H}_1}$, in accordance with the set $\{\cket{\alpha}\}_\alpha$, and choose $\rho$ and $\oper{W}$ accordingly.} of identity map $\mathcal{I}=\sum_\alpha \cketbra{\alpha}{\alpha}$.

Consequently, Eq.~\eqref{eq:w_t-gen} reads
		\begin{equation}
			w_{\rho}\cv{t} = \sum_{\alpha}e^{\lambda_\alpha t}\big(\rho\big\vert\Big[\big\vert \alpha)\big(\alpha\big\vert\otimes\mathcal{I}\Big] \big\vert\oper{W}\big)\label{eq:w_t-decom}.
		\end{equation}
Let us rewrite \[\big(\rho\big\vert\Big[\big\vert \alpha)\big(\alpha\big\vert\otimes\mathcal{I}\Big] \big\vert\oper{W}\big)=r_\alpha e^{i\phi_\alpha},\]
which is just a polar decomposition of a complex number, and denote
$\Gamma_\alpha=\Re\lambda_\alpha\leqslant 0$ and $\omega_\alpha=\Im\lambda_\alpha$ to respectively be the decay rates and oscillation frequencies associated with the dynamics. In the Appendix we show that all these parameters split into three classes, summarized in Table~\ref{tab:classification}. Looking at the table one can immediately recognize that the function $w_{\rho}\cv{t}$ is real, so that
\begin{equation}
    \Im w_{\rho}\cv{t} = \sum_{\alpha}r_\alpha e^{\Gamma_\alpha t}\sin\cv{\omega_\alpha t+\phi_\alpha} =0,
\end{equation} and Eq.~\eqref{eq:w_t-decom} is equivalent to
	\begin{equation}
		w_{\rho}\cv{t}=\sum_{\alpha}r_\alpha e^{\Gamma_\alpha t}\cos\cv{\omega_\alpha t+\phi_\alpha} \label{eq:w_t_real}.
	\end{equation}
We observe that the negative contributions, which reflect entanglement, are associated with oscillation frequencies $\omega_\alpha$ and  initial phases $\phi_\alpha$. The first type of parameters describes the sole properties of the dynamics, while the latter type refers to the relative direction of the dynamical axes with respect to prepare-measurement configuration. 

\subsection{Structure of the Dynamical Components}\label{sec:struct}
From the previous discussion we can see that one can characterize the behavior of the witness looking at the interplay between different parameters involved. In particular, one can decompose the average of the witness as follows:\begin{subequations}
	\begin{equation}
		w_{\rho}\cv{t} = w_I + w_{II}\cv{t} + w_{III}\cv{t} \label{eq:w_split},
	\end{equation}
where:
	\begin{align}
		w_I &= \big(\rho\big\vert\Big[\big\vert  0\big)\big( 0\big\vert\otimes\mathcal{I}\Big] \big\vert\oper{W}\big)=r_0 e^{i \phi_0}=\pm r_0\label{eq:W_I},\\
		w_{II}\cv{t} &=\sum_{j=1}^{K}r_j e^{\Gamma_j t}\cos\cv{\phi_j}, ~~\phi_j\in\{0,\pi\} \label{eq:W_II},\\
		w_{III}\cv{t} &= 2\sum_{j=1}^{L}\tilde{r}_j e^{\tilde{\Gamma}_j t}\cos\cv{\tilde{\omega}_jt + \tilde{\phi}_j}\label{eq:W_III}.
	\end{align}
Clearly $w_I$ is the constant term corresponding to $\lambda_0=0$, while $w_{II}(t)$ represents the decay.  
\end{subequations}

We relabeled $\Gamma_{K+1},\ldots,\Gamma_{d}$ as $\tilde{\Gamma}_{1},\ldots,\tilde{\Gamma}_{2L}$, where $L$ is the number of elements in the class $III$.  The same pattern has been applied to $\tilde{\omega}_j,$ $\tilde{r}_{j},$ and $\tilde{\phi}_{j}$. Note that the total dimension decomposes as $L=\cv{d^2-K-1}/2$. Note also that the eigenvector associated with $\lambda_0=0$, an identity operator $\iden$, which constitutes the class $I$, is denoted by a vector $\cket{0}$. 

Let us now remark on an interesting case $w_{I}=r_0\geqslant 0$, corresponding to $\phi_0=0$. This can occur, for example, when
\[\Big[\big\vert  0)\big( 0\big\vert\otimes\mathcal{I}\Big] \big\vert\oper{W}\big) \propto \cket{0}^{\otimes 2},\]
or  when $\oper{W}$ is orthogonal to the identity. In this scenario the term  $w_{II}\cv{t}+w_{III}\cv{t}$ will contribute to the negativity of  $w_{\rho}\cv{t}$ while the time-independent term $w_{I}$ will set the threshold for the former terms. In other words, entanglement remains at time $t$ if
	\begin{equation}
		\cvv{w_{II}\cv{t}+w_{III}\cv{t}} > w_I. \label{eq:ent_cri}
	\end{equation}

%good config
Let us also point out a specific setting, in which one prepares a pair $\cv{\rho,\oper{W}}$ in such a way that $\phi_j=\tilde{\phi}_j=\pi$ for $j>0$. At the initial time $t=0$ the summands in Eqs.~\eqref{eq:W_II} and \eqref{eq:W_III} are all negative. Furthermore, we also observe that at time $t\geqslant 0$
	\begin{align}
		w_{II}\cv{t} &= -\sum_{j=1}^{K}r_j e^{\Gamma_j t} \leqslant 0\label{eq:W_II-opt},\\
		w_{III}\cv{t} &= -2\sum_{j=1}^{L}\tilde{r}_j e^{\tilde{\Gamma}_j t}\cos\cv{\tilde{\omega}_jt} \label{eq:W_III-opt}.
	\end{align}
The term $w_{II}\cv{t}$ is  an increasing function eventually approaching $0$. We call such special prepare-measure pairs of operators $\cv{\rho,\oper{W}}$ \textit{good configurations}. Given that a good configuration exists for every channel (which indeed is the case, as constructively shown in Sec. \ref{sec:W-Psi}), we infer that coherent dynamics speeds up the deterioration of the EP property.

In the following section we derive bounds for $t_{CB}$ using two standard methods relevant for quantum speed limit and mentioned in the introduction, as well as obtain a specific bound valid for good configurations. While the  Mandelstam–Tamm-inspired bound will be valid in general, the  Margolus–Levitin-inspired bound will also only apply to good configurations.

\section{Entanglement Breaking Time}\label{sec:Breaking-time}
%say that it is hard in general
In general, the problem of finding $t_{CB}\cv{\rho,\oper{W}}$ defined in Eq.~\eqref{eq:t_B-finding}, with explicit form of the input function taken from Eq.~\eqref{eq:w_t_real}, is very complicated.  This is due to its non-linear character and plenty of involved parameters. In fact, since $w_\rho\cv{0} < 0$, this problem boils down to solving highly nonlinear equation $w_\rho\cv{t} = 0$, and further seeking for the minimum root. Obviously, if all parameters are known, one can employ numerical methods to find the exact breaking time. Therefore, our goal is to get explicit results without specifying the parameters. To this end we shall follow three routes in order to bound $t_{EB}$. 

\subsection{Mandelstam–Tamm-inspired Bound}\label{sec:generic}
%general bound for general witness
In this part we consider the general case pertaining to an arbitrary prepare-measure pair $\cv{\rho,\oper{W}}$. 

Firstly we follow the most standard approach towards quantum speed limit. We observe that 
\begin{align}
    w_{\rho}&\left(t\right)-w_{\rho}\left(0\right)  =  \int_{0}^{t}dt\dot{w}_{\rho}\left(t\right) =  \int_{0}^{t}dt\sum_{\alpha}r_{\alpha}\lambda_\alpha e^{\lambda_{\alpha}t +i\phi_\alpha} \label{eq:w-gen-int}.
\end{align}
Since $r_{\alpha}\geqslant 0$ and $\left|e^{\lambda_{\alpha}t +i\phi_\alpha}\right|=\left|e^{\Gamma_{\alpha}t}\right|\leqslant 1$, we are able to bound
\begin{eqnarray}
    \left\vert w_{\rho}\left(t\right)-w_{\rho}\left(0\right) \right\vert & = &  \left\vert\int_{0}^{t}dt\sum_{\alpha}r_{\alpha}\lambda_\alpha e^{\lambda_{\alpha}t +i\phi_\alpha}\right\vert \nonumber\\
    &\leqslant&\int_{0}^{t}dt\sum_{\alpha}r_{\alpha}\left\vert\lambda_\alpha\right\vert\nonumber\\
    & =&t \sum_{\alpha}r_{\alpha}\left\vert\lambda_{\alpha}\right\vert,
\end{eqnarray}
where the last equation just follows from evaluating the remaining trivial integral. Consequently, we get
\begin{equation}
t\geqslant\dfrac{\left\vert w_{\rho}\left(t\right)-w_{\rho}\left(0\right)\right\vert}{\sum_{\alpha}r_{\alpha}\left\vert\lambda_{\alpha}\right\vert} \label{eq:t-bound-gen}.
\end{equation}
Note that $\left\vert\lambda_{\alpha}\right\vert=\sqrt{\Gamma_{\alpha}^2+\omega_{\alpha}^2}.$ Since $w_{\rho}\left(0\right)<0$ and because $w_\rho\cv{t_{CB}}=0$, we get the first final result
\begin{equation}
t_{CB}\geqslant \dfrac{\left\vert w_{\rho}\left(0\right)\right\vert}{\sum_{\alpha}r_{\alpha}\left\vert\lambda_{\alpha}\right\vert}. \label{eq:t_B-gen}
\end{equation}
Note that
\begin{equation}
w_{\rho}\left(0\right)=\sum_{\alpha}r_{\alpha}\cos\left(\phi_{\alpha}\right). \end{equation}

%discuss the W and rho
As already mentioned, the time $t_{CB}$ also depends on the details of the prepare-measure configuration $\cv{\rho,\oper{W}}$. From now on we intend to leverage our results by appropriately using this freedom. Therefore, all remaining results of this section will be derived under an assumption that the pair $\cv{\rho,\oper{W}}$ forms a good configuration. Then, in  Sec. \ref{sec:W-Psi} we explicitly construct such a configuration, in which the state is maximally entangled while the witness is based on the projection on this very special state.

\add{On a first sight, one could attempt to use the same approach in order to describe a future time moment in which a potential revival of EP takes place. Such \textit{configuration EP revival} time $t_{CREP}>t_{CB},$ if it exists, would then need to be defined according to the condition $w_{\rho}\left(t_{CREP}\right)=-\epsilon$, for any $\epsilon>0$. Consequently,
\begin{equation}
    t_{CREP}-t_{CB}\geqslant \dfrac{\left\vert \epsilon\right\vert}{\sum_{\alpha}r_{\alpha}\left\vert\lambda_{\alpha}\right\vert}. 
\end{equation}
We immediately see that the bound assumes an infinitesimal value, so it is not informative. One could still try to modify the above procedure; however, it seems clear that the revival of EP will not be captured in a similar fashion to the EB property.}

\subsection{General bound for Good Configurations}\label{sec:eff-bound}
%recall a bit and propose the way to make it better
Let $\Gamma_l$ be a lower bound for all real parts of the nontrivial eigenvalues $\{\gamma_\alpha\}_{\alpha\neq 0},$ i.e., $\Gamma_l\leqslant\Gamma_\alpha$ for all $\alpha$. Because of Eq.~\eqref{eq:W_II-opt}, fulfilled by good prepare and measure configurations, we can provide a bound
\begin{equation}
    	w_{II}\cv{t} \leqslant -e^{\Gamma_l t}C_{II} \leqslant 0, \label{eq:bounds_w_II}
\end{equation}
where $C_{II}=\sum_{j=1}^{K}r_j$. In a similar fashion, if we denote
\begin{equation}
    \tilde\Omega=\max_j \tilde\omega_j,
\end{equation}
we get
\begin{equation}
     w_{III}\cv{t} \leqslant -e^{\Gamma_l t}C_{III}(T) \leqslant0, \label{eq:bounds_w_III}
\end{equation}
where  $C_{III}(T)=2\sum_{j=1}^{L}\tilde{r}_j\cos\cv{\tilde{\omega}_jT}$,
valid for $0\leqslant t\leqslant T$ whenever $T\leqslant \pi/(2\tilde\Omega)$. This is true because $\cos\cv{\tilde{\omega}_jT}\geqslant0$ for all $j$. Consequently, if $t_{CB}\leqslant T$, after a few algebraic steps we get the lower bound
    \begin{equation}
        t_{CB} \geqslant  \dfrac{1}{\cvv{\Gamma_l}}\ln\Bigg(\dfrac{C_{II}+C_{III}(T)}{w_I}\Bigg):=\tau_{CB}(T). \label{eq:w-opt-summary}
    \end{equation}
By the above arguments we get that if the right-hand side of the last inequality is smaller than $T$, it forms a valid lower bound on $t_{CB}$. On the other hand, if this is not the case, we know that $t_{CB}$ is at least $T$. Therefore, for any $0\leqslant T\leqslant \pi/(2\tilde\Omega)$ we find that 
\begin{equation}
        t_{CB} \geqslant  \min\left\{T,\tau_{CB}(T)\right\}.
    \end{equation}
Consequently, we also get
\begin{equation}\label{maximizing}
        t_{CB} \geqslant \max_{0\leqslant T\leqslant \pi/(2\tilde\Omega)} \min\left\{T,\tau_{CB}(T)\right\}.
    \end{equation}
We stress that, contrarily to the general result in Eq. (\ref{eq:t_B-gen}) the above bound only holds for good prepare-measure configurations. In particular, the logarithm therein requires that $w_I>0$, which is one of the characteristic features of such configurations. 

\subsection{Margolus–Levitin-inspired Bound for Good Configurations}
For the sake of completeness we shall also discuss a variant of the bound inspired by the Margolus–Levitin bound. However, the method used to derive this bound turns out to be unsuitable in the general case. This happens because that bound is based on the inequality 
\begin{equation}\label{cosb}
    \cos\cv{x} \geqslant 1-\dfrac{2}{\pi}\left[x+\sin\cv{x}\right],
\end{equation} valid for $x\geqslant 0$. In our problem, this inequality could potentially be applied to bound $\cos(\omega_\alpha t+\phi_\alpha)$ factors in (\ref{eq:w_t_real}). However, Eq. (\ref{cosb}) constitutes a lower bound which, given that we rely on $w_\rho(t_{CB})\geqslant0$, is not useful. Moreover, in general we have no control over the sign of the arguments  $\omega_\alpha t+\phi_\alpha$.

Quite interestingly, both limiting factors pointed out above disappear for good configurations. First of all, the trigonometric terms appear only in $w_{III}(t)$ and are always multiplied by $-1$. Moreover, since all $\tilde\omega_j$ are nonnegative, we also do not suffer from the sign issue. 

Therefore, for good configurations we can bound $w_{III}(t)$ as follows:
\begin{eqnarray}
w_{III}(t) & = & -2\sum_{j=1}^{L}\tilde{r}_{j}e^{\tilde{\Gamma}_{j}t}\cos\left(\tilde{\omega}_{j}t\right)\\
 & \leqslant & 2\sum_{j=1}^{L}\tilde{r}_{j}e^{\tilde{\Gamma}_{j}t}\left[\frac{2}{\pi}\tilde{\omega}_{j}t+\frac{2}{\pi}\sin\left(\tilde{\omega}_{j}t\right)-1\right]\nonumber\\
 & \leqslant & 2\sum_{j=1}^{L}\tilde{r}_{j}e^{\tilde{\Gamma}_{j}t}\left[\frac{2}{\pi}\tilde{\omega}_{j}t-\frac{\pi-2}{\pi}\right].\nonumber
\end{eqnarray}
In the last line we just bounded the sin function by $1$. Since all
$\tilde{\omega}_{j}\geqslant 0$, the first term is positive so that the
exponent multiplying it can be bounded by 1 (since all $\tilde{\Gamma}_{j}\leqslant 0$).
On the other hand, the second term is negative, so we can bound the
exponent as follows:
\begin{equation}
e^{\tilde{\Gamma}_{j}t}=e^{-\left|\tilde{\Gamma}_{j}\right|t}\geqslant 1-\left|\tilde{\Gamma}_{j}\right|t,
\end{equation}
using the fact that $e^{-x}\geqslant 1-x$ for $x\geqslant0$. As a result we
get the inequality
\begin{equation}
w_{III}(t)\leqslant 2\sum_{j=1}^{L}\tilde{r}_{j}\left[\frac{2}{\pi}\tilde{\omega}_{j}t+\frac{\pi-2}{\pi}\left(\left|\tilde{\Gamma}_{j}\right|t-1\right)\right].
\end{equation}
Following the same reasoning concerning the exponential decay, we also
bound
\begin{equation}
w_{II}(t)\leqslant\sum_{j=1}^{K}r_{j}\left(\left|\Gamma_{j}\right|t-1\right).
\end{equation}
Finally, since $w_{\rho}\left(t_{CB}\right)\geqslant 0$, we get the lower
bound
\begin{equation}
t_{CB}\geqslant\frac{\sum_{j=1}^{K}r_{j}+2\frac{\pi-2}{\pi}\sum_{j=1}^{L}\tilde{r}_{j}-r_{0}}{\sum_{j=1}^{K}r_{j}\left|\Gamma_{j}\right|+2\sum_{j=1}^{L}\tilde{r}_{j}\left(\frac{2}{\pi}\tilde{\omega}_{j}+\frac{\pi-2}{\pi}\left|\tilde{\Gamma}_{j}\right|\right)}.
\end{equation}

The above bound for the entanglement breaking time looks rather cumbersome. It does not only depend on the spectrum of the channel (superoperator) represented by the parameters $\Gamma_j$, $\tilde \Gamma_j$, and $\tilde\omega_j$, but also on an interplay between a state, an entanglement witness and the eigenbasis of the superoperator (through $r_0$, $r_j$, and $\tilde r_j$). In the next section we select both the state and the entanglement witness in such a way that together they not only form a good configuration but also, due to very high symmetry of this configuration, render the parameters ``r'' which do not depend on the basis $\big\vert \alpha)$.

\section{Symmetric Entanglement Witness}\label{sec:W-Psi}
In this section we consider a specific choice for the entanglement witness operator
%\com{FS:I change it back.}
\begin{equation}
\hat{W}_{\Psi_+}=\iden^{\otimes 2}-d\left\vert\Psi_{+}\right\rangle \left\langle \Psi_{+}\right\vert \label{eq:W_d},
\end{equation}
with the maximally entangled state $\left\vert \Psi_{+}\right\rangle$ already defined in (\ref{MaxEntS}). One can observe that the average value of this witness is $0$ for all separable states, a fact which suggests a certain optimality of this choice of the witness.

Moreover, in order to strengthen the configuration we also select the state to be maximally entangled, i.e. $\hat \rho=\left\vert\Psi_{+}\right\rangle \left\langle \Psi_{+}\right\vert$. We shall call this whole choice a \textit{symmetric configuration}. 

As a result, the time-dependent average value of the witness becomes
\begin{align}
    w_{\Psi_+}\left(t\right)  &= 1 - d\big(\Psi_+\big\vert\Phi_t\otimes\mathcal{I} \big\vert\Psi_+\big),\nonumber\\
        &=1- d\sum_{\alpha}e^{\lambda_\alpha t}s_\alpha \label{eq:w_sym},
\end{align}
where
\begin{equation}
s_\alpha=\big(\Psi_+\big\vert\Big[\big\vert \alpha)\big(\alpha\big\vert\otimes\mathcal{I}\Big] \big\vert\Psi_+\big). \label{eq:coeff_s}
\end{equation}
In the vectorized notation, the state
\begin{equation}
    \hat \rho=\left\vert\Psi_{+}\right\rangle \left\langle \Psi_{+}\right\vert=\dfrac{1}{d}\sum_{j,j'=1}^d\ketbra{j}{j'}\otimes\ketbra{j}{j'},
\end{equation}
is represented by a vector
\begin{equation}
        \cket{\Psi_{+}} = \dfrac{1}{d}\sum_{j,j'=1}^d\cket{e_{jj'}}\otimes\cket{e_{jj'}}, \label{eq:Psi}
    \end{equation}
where $\cket{e_{jj'}}$ are members of a canonical basis of the matrix space $\mathcal{B}\cv{\mathcal{H}_1}$. In other words, $\cket{e_{jj'}}$ is a vectorized form of $\ketbra{j}{j'}$ (we remember that $\mathcal{H}_1=\mathcal{H}_2\simeq \mathbb{C}^d$).

We are in position to use the above vectorization to prove two technical results concerning the choice (\ref{eq:w_sym}):
\begin{lemma}\label{Lemma1}
For every channel $\Phi_t$ we have that
\begin{equation}
\forall_{\alpha}\quad s_{\alpha}=\frac{1}{d^{2}}.
\end{equation}
\end{lemma}
To show this result, one shall perform the calculation as follows (we omit summation ranges for brevity):
    \begin{eqnarray}
         s_\alpha &=& \cbra{\Psi_+}\left[\cketbra{\alpha}{\alpha}\otimes\mathcal{I}\right]\cket{\Psi_+}\nn\\
            &=& \dfrac{1}{d^2}\sum_{j,j',j'',j'''}\cbraket{e_{jj'}}{\alpha}\cbraket{\alpha}{e_{j''j'''}}\cbraket{e_{jj'}}{e_{j''j'''}}\nn\\
      &=& \dfrac{1}{d^2}\sum_{j,j'}\cbraket{\alpha}{e_{jj'}}\cbraket{e_{jj'}}{\alpha}= \dfrac{1}{d^2}\cbraket{\alpha}{\alpha} = \dfrac{1}{d^2} \label{eq:s_alpha}.
    \end{eqnarray}
Passing from the second to the third line we used orthogonality of the vectors $\cket{e_{jj'}}$.
\begin{lemma}\label{Lemma2}
For every channel $\Phi_t$ we have that
\begin{align}
         r_\alpha e^{i\phi_\alpha} &= \delta_{\alpha 0} - d s_\alpha=\left\{\begin{array}{lr}
            1-\frac{1}{d},  &  \alpha=0\\
            -\dfrac{1}{d},  & \text{~otherwise}
         \end{array} \right. \label{eq:coef_sym},
    \end{align}
    provided that $\cv{\rho,\oper{W}}$ is a symmetric configuration.
\end{lemma}
First of all, since the witness under discussion is vectorized to the form $\big\vert\oper{W}\big)=\big\vert0\big)^{\otimes 2}-d \cket{\Psi_{+}}$, for the symmetric configuration we know that
    \begin{align*}
        r_\alpha &e^{i\phi_\alpha} = \big(\rho\vert\Big[\big\vert \alpha)\big(\alpha\big\vert\otimes\mathcal{I}\Big] \big\vert\oper{W}\big)\\
            &= \big(\Psi_+\big\vert\Big[\big\vert \alpha)\big(\alpha\big\vert\otimes\mathcal{I}\Big] \big\vert0\big)^{\otimes 2}- d\big(\Psi_+\big\vert\Big[\big\vert \alpha)\big(\alpha\big\vert\otimes\mathcal{I}\Big] \big\vert\Psi_+\big)\\
            &= \big(\Psi_+\big\vert\Big[\big\vert \alpha)\big(\alpha\big\vert\otimes\mathcal{I}\Big] \big\vert0\big)^{\otimes 2} - d s_\alpha.
    \end{align*}
Therefore, given Lemma \ref{Lemma1} we only need to explicitly calculate the first term. To this end, we observe that $\cket{0}$, which corresponds to the identity operator $\iden=\sum_{j=1}^d \ketbra{j}{j}$, is represented as $\cket{0}=\sum_{j=1}^d\cket{e_{jj}}$. We can then explicitly calculate
    \begin{align}
       \hspace{1cm}&\hspace{-1cm}\big(\Psi_+\big\vert\Big[\big\vert \alpha)\big(\alpha\big\vert\otimes\mathcal{I}\Big] \big\vert0\big)^{\otimes 2}\nn\\
            &= \dfrac{1}{d}\sum_{j,j'}\cbraket{e_{jj'}}{\alpha}\cbraket{\alpha}{ 0}\cbraket{e_{jj'}}{ 0}\nn\\
            &= \dfrac{1}{d}\sum_{j,j'}\cbraket{e_{jj'}}{\alpha}\delta_{\alpha 0}\Big[\sum_{j''}\cbraket{e_{jj'}}{e_{j''j''}}\Big]\nn\\
            &= \dfrac{\delta_{\alpha 0}}{d}\sum_{j,j'}\cbraket{e_{jj'}}{\alpha}\sum_{j''}\delta_{jj''}\delta_{j'j''}\nn\\
            &= \dfrac{\delta_{\alpha 0}}{d}\sum_{j,j'}\cbraket{e_{jj'}}{\alpha}\delta_{jj'}\nn\\
            &= \dfrac{\delta_{\alpha 0}}{d}\sum_{j}\cbraket{e_{jj}}{ 0}= \dfrac{\delta_{\alpha 0}}{d}\sum_{jj'}\cbraket{e_{jj}}{e_{j'j'}}=\delta_{\alpha 0} \label{eq:coef_sym_fixed-term}.
    \end{align}
This finalizes the proof. Given both lemmas above, we reach the following conclusion
\begin{corollary}
    The symmetric configuration is also a good configuration.
\end{corollary}
Lemma \ref{Lemma2} says that $r_0=1-\frac{1}{d}\geqslant0$ and consequently $\phi_0=0$. On the  other hand, for $\alpha\neq0$ (i.e., for members of classes II and III) we can see that $r_\alpha e^{i\phi_\alpha}$ is real and negative. Therefore, $\phi_\alpha=\pi$ in all these cases. These are exactly the conditions defining the good configuration.
    
As we can see, one can find a good configuration for any channel $\Phi_t$, simply by means of the symmetric configuration discussed in this section. Moreover, since the entanglement breaking time $t_{EB}$ is lower bounded by all $t_{CB}$, we conclude that all three bounds derived in the previous section do apply to $t_{EB}$. 

In the following, we simplify these bounds given the symmetric configuration. Before doing so, we note in passing that the symmetric configuration has an additional interesting feature, namely, it can be related to the geometric measure of entanglement \cite{GEOM}. The latter was recently shown to be equal to a minimal time required for a unitary (global) transformation to transform a given pure entangled state to a closed separable state~\cite{Rudnicki2021}. An analog of $t_{EB}$ in this problem reads $\Omega^{-1}\arccos\left(\sqrt{d}\right)$ with $\Omega$ defined as an energy scale of a Hamiltonian  rendering the global time evolution. Recently, a similar problem has been studied in Refs.~\cite{Das2019,Das2022}, where the dynamical behavior of several entanglement quantifiers has been considered for a generic form of dynamics, revealing a relation between quantum speed limit and the change in entanglement.

\subsection{Speed of entanglement breaking property}\label{summaryofbounds}
We are now going to summarize the above findings by bounding time when the channel becomes EB as
\begin{equation}
    t_{EB}\geqslant t_{CB}\cv{\left\vert\Psi_{+}\right\rangle \left\langle \Psi_{+}\right\vert,\oper{W}_{\Psi_+}}.
    \end{equation}
The Mandelstam-Tamm-inspired lower bound for the symmetric configuration provides
\begin{equation}
    t_{EB}\geqslant \dfrac{d\cv{d-1}}{\sum_{\alpha}\left\vert\lambda_{\alpha}\right\vert}=T_{\mathrm{M-T}} \label{eq:t_B-gen-sym}.
\end{equation}
In the case of the general bound for good prepare-measure configurations we can slightly simplify the intermediate bound defined in (\ref{eq:w-opt-summary}) to the form
    \begin{equation}
        \tau_{CB}(T)=\dfrac{1}{\cvv{\Gamma_l}}\ln\Bigg(\dfrac{K+2\sum_{j=1}^{L}\cos\cv{\tilde{\omega}_jT}}{d-1}\Bigg). \label{eq:w-opt-summary-sym}
    \end{equation}
Still, this bound does depend on the time threshold $T$, and needs to be optimized as in (\ref{maximizing}). For consistency, let us denote such an optimized bound by $T_{GC}$. Finally, The Margolus–Levitin-inspired  lower bound for the symmetric configuration gives
\begin{equation}
t_{EB}\geqslant\frac{K+2\frac{\pi-2}{\pi}L-d+1}{\sum_{j=1}^{K}\left|\Gamma_{j}\right|+2\sum_{j=1}^{L}\left(\frac{2}{\pi}\tilde{\omega}_{j}+\frac{\pi-2}{\pi}\left|\tilde{\Gamma}_{j}\right|\right)}=T_{\mathrm{M-L}}.
\end{equation}

\subsection{The Qubit case}\label{sec:qubit-eg}
\begin{figure}[!ht]
\begin{flushleft}
a)
    \includegraphics[width=\linewidth]{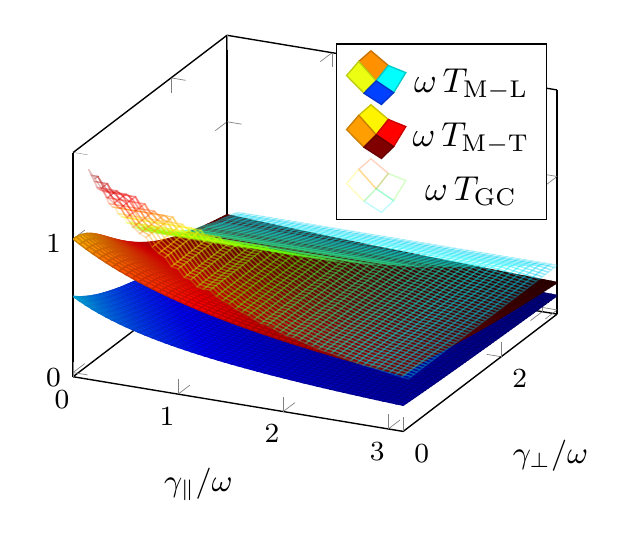}
b)
    \includegraphics[width=\linewidth]{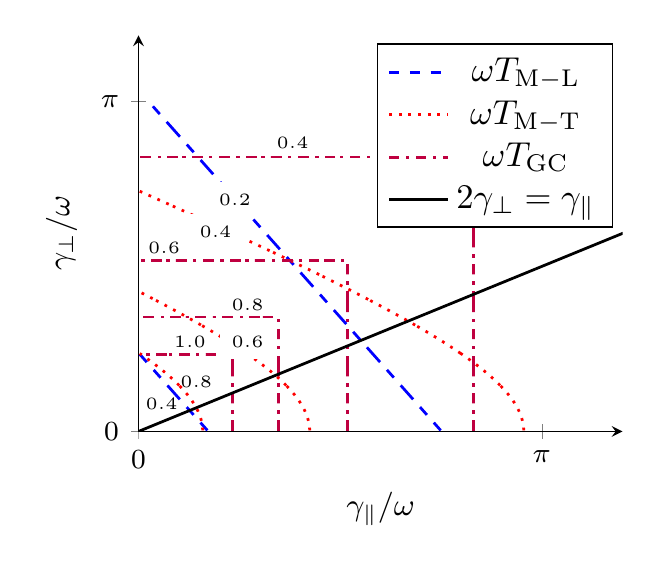}
    \caption{Comparisons of $T_{\mathrm{M-T}},$ $T_{\mathrm{GC}}$ and $T_{\mathrm{M-L}}$ for qubit dynamical map given in Eq. \eqref{eq:eg-qubit} for different values of $\gamma_\perp$ and $\gamma_\parallel$ modulated by $\omega$ 
    \add{(therefore all quantities are given in dimensionless units)}. The contours of the functions in (a) are shown in (b), labeled by their corresponding values $T_{\mathrm{M-T}},$ $T_{\mathrm{GC}}$ and $T_{\mathrm{M-L}}$ [their heights in (a) respectively]. The black thick line represents the physical condition for complete positivity of the dynamical map, i.e., only the maps with parameters above such line are physically allowed. One can observe that, in the physical regions of parameters we consider, the bounds satisfy the order $T_{\mathrm{GC}}>T_{\mathrm{M-T}}>T_{\mathrm{M-L}}.$}
    \label{fig:T-qubit}
\end{flushleft}
\end{figure}
In this section we demonstrate implications of derived bounds via an example of two qubits, (i.e. $d=2$) undergoing local and unital Lindblad dynamics. Without loss of generality, we consider the evolution map in the Heisenberg picture given by
    \begin{equation}
        \Phi^*_t \equiv \text{diag}\cv{1,e^{-\gamma_\parallel t},e^{-\gamma_\perp t}e^{-it\omega},e^{-\gamma_\perp t}e^{it\omega}} \label{eq:eg-qubit},
    \end{equation}
where $\gamma_\parallel,\gamma_\perp \geqslant 0$ and $\pi\geqslant\omega\geqslant 0$. These parameters need to satisfy the complete positivity condition $2\gamma_\perp\geqslant\gamma_\parallel$ \cite{Chruscinski2021}. The right-hand side of (\ref{eq:eg-qubit}) gives the matrix representation of $\Phi^*_t$ in the $\cket{\alpha}$ basis, for $\alpha=0,\ldots,3$. Although the above expression represents the spectral decomposition in Eq.~\eqref{eq:Phi-decom}, $\gamma_\parallel=-\cvv{\Gamma_1}$, $\gamma_\perp=-\cvv{\tilde{\Gamma}_1}$ and $\omega=\tilde{\omega}_1$, these three parameters at the same time correspond to the characteristics of the Markovian qubit dynamics. In other words, $\gamma_\parallel$ and $\gamma_\perp$ are decay rates in longitudinal and transversal degrees of freedom with respect to a quantization axis set by the system, while $\omega$ is the precession frequency about such axis \cite{Alicki2007}. Note that, in general, the parameter $\omega$ can also be $0$, but we only consider nonzero values of $\omega$ for clarity. 

The two explicit bounds established in this paper become
    \begin{align}
        T_{\mathrm{M-T}}&\cv{\omega,\gamma_\parallel,\gamma_\perp}=\dfrac{2}{\gamma_\parallel +2\sqrt{\omega^2+\gamma_\perp^2}}\label{eq:T_TM-qubit},\\
        T_{\mathrm{M-L}}&\cv{\omega,\gamma_\parallel,\gamma_\perp}=\dfrac{\pi-2}{ \tfrac{\pi}{2}\gamma_\parallel + 2\omega +\cv{\pi-2}\gamma_\perp}
        \label{eq:T_ML-qubit}.
    \end{align}
Optimization rendering the bound $ T_{\mathrm{GC}}\cv{\omega,\gamma_\parallel,\gamma_\perp}$ leads to a transcendental equation
\begin{equation}
    T_{GC}=\dfrac{\ln\cv{1+2\cos\cv{\omega T_{GC}}}}{\max\cvc{\gamma_\parallel,\gamma_\perp}}, \label{eq:opt-tau_T-qubit}
\end{equation}
just because the right-hand side of (\ref{eq:w-opt-summary-sym}) becomes a decreasing function of $T$ for $0\leq T \leq \pi/2\omega$. 

Let us now compare these three bounds regarding different regions of the parameters $\cv{\gamma_\parallel/\omega,\gamma_\perp/\omega},$ where the modulation by the frequency is used for simplicity. 
Figure~(\ref{fig:T-qubit}a) suggests  that $ T_{\mathrm{GC}}>T_{\mathrm{M-T}}>T_{\mathrm{M-L}}$ for all physically meaningful values of the parameters. In fact, there is a region where $T_{\mathrm{GC}}<T_{\mathrm{M-T}}$ but its associate parameters $\gamma_\parallel$ and $\gamma_\perp$ do not satisfy complete positivity condition $2\gamma_\perp\geqslant\gamma_\parallel$ --- see Fig.~\ref{fig:T-qubit}(b). In other words, for the qubit case we observe that the dynamical map $\Phi_t$ is certainly entanglement preserving for all times $t<T_{\mathrm{GC}}\cv{\omega,\gamma_\parallel,\gamma_\perp}$.
We believe that this feature is specific to qubits, since in general one could expect different orders among $T_{\mathrm{M-T}}$, $T_{\mathrm{GC}}$, and $T_{\mathrm{M-L}}$ for different regions of the dynamical parameters $\cvc{\omega_\alpha}$ and $\cvc{\Gamma_\alpha},$ subject to complete positivity. 
%%%%%%%%%%%%%%%%%%%%%%%%%%%%%%%%%%%%%%%%%%%%%%%%%%%%%%
%%%% DISCUSSION AND CONCLUSIONS	
%%%%%%%%%%%%%%%%%%%%%%%%%%%%%%%%%%%%%%%%%%%%%%%%%%%%%%
\section{Conclusion}\label{sec:conclusion}
Since time in quantum mechanics is not an observable --- it is ``just" a parameter --- the rich formalism of uncertainty relations cannot be utilized to describe the dynamics. However, quantum speed limit, even though based on slightly different foundations, often seems to offer a sufficient quantification of various dynamical features of quantum systems. Here we propose yet another aspect in which the methodology behind the QSL can successfully be applied. We study Markovian open-system dynamics which, when described in the language of quantum channels, always corresponds to an identity channel in the starting moment of the evolution. Consequently, such a dynamics always starts with an entanglement-preserving channel. Therefore, depending on the details of the Markovian dynamics, such a channel sooner (but not instantly) or later (perhaps never) becomes entanglement breaking. While the task to describe this transition exactly is \add{computationally laborious} as it would pretend to solving an NP-hard problem \cite{Gharibian}, using the QSL we managed to derive three lower bounds  for the time moment in which it happens (see Sec. \ref{summaryofbounds} for a summary). These bounds do only depend on the parameters describing  the associated master equation, so that they can be expressed in terms of decay rates and oscillation frequencies. Despite fundamental aspects pertaining to a better understanding of quantum open-system dynamics, the presented results in a way complement other efforts aiming at description of generation and degradation of quantum resources \cite{KavanNJP}.

Future, more technically oriented studies can already start from the case of qutrits ($d=3$). Even for a specific map similar to \eqref{eq:eg-qubit}, namely,
    \begin{equation}
        \Phi^*_t \equiv \text{diag}\cv{1}\bigoplus_{k=1}^4 \text{diag}\cv{e^{-\gamma_k t}e^{-it\omega_k},e^{-\gamma_k t}e^{it\omega_k}} \label{eq:eg-qutrit},
    \end{equation}
where all parameters are defined in the same spirit as $\gamma_\parallel$, $\gamma_\perp$, and $\omega$ in the qubit case, one can see that there are four pairs of real parameters to be considered. This setting immediately leads to more complex relations among the obtained bounds. We shall leave this example, as well as more elaborate cases, as an open question for further development.

\section*{Acknowledgments}
We thank Kavan Modi for a stimulating discussion and Sevag Gharibian for fruitful correspondence. We acknowledge support by the Foundation for Polish Science (IRAP project, ICTQT, Contract No. 2018/MAB/5, co-financed by EU within the Smart Growth Operational Programme).

\appendix
\section*{APPENDIX: Absence of Imaginary Part of the Entanglement Witness}\label{appen:no-Im-w}
In the main text we split the parameters into three different classes and for the sake of brevity just observed that this fact easily implies that $w_\rho(t)$ is real. Here, we aim to show the latter result in an independent way. As a by-product we shall find that this is equivalent to the content of Table \ref{tab:classification}.

In other words, here we explain why the imaginary part $\sum_{\alpha}r_\alpha e^{\Gamma_\alpha t}\sin\cv{\omega_\alpha t+\phi_\alpha}$ vanishes. Recall Eq.~\eqref{eq:w_t-decom}
	\begin{equation*}
			w_{\rho}\cv{t} = \sum_{\alpha}e^{\lambda_\alpha t}\big(\rho\big\vert\Big[\big\vert \alpha)\big(\alpha\big\vert\otimes\mathcal{I}\Big] \big\vert\oper{W}\big).
	\end{equation*}
We first notice that $\mathcal{L}^*$ is Hermiticity preserving, i.e., $\mathcal{L}^*\cv{X^\dagger}=\mathcal{L}^*\cv{X}^\dagger,$ since the operation $\Phi^*_t$ is completely positive. This leads to $\mathcal{L}^*\cket{\alpha^\dagger}=\lambda_{\alpha^\dagger}\cket{\alpha^\dagger}=\overline{\lambda}_\alpha\cket{\alpha^\dagger}$ being also another eigen equation where $\cket{\alpha^\dagger}$ is a Hermitian conjugate of $\cket{\alpha}.$
The dual element of $\cket{\alpha^\dagger}$ appears to be the conjugate of $\cket{\alpha},$ i.e., $\cket{\alpha^\dagger}=\cket{\alpha}^\dagger.$ This comes from
    	\[	\delta_{\alpha\alpha'}=\cbraket{\alpha}{\alpha'}=\overline{\cbraket{\alpha}{\alpha'}}=\cbraket{\alpha^\dagger}{\alpha'^\dagger}, \]
supplied by invariance  of the trace (defining the above inner product) with respect to transposition. From this relation, it follows that 
	\begin{equation}
    	\big(\rho\big\vert\Big[\big\vert \alpha^\dagger)\big(\alpha^\dagger\big\vert\otimes\mathcal{I}\Big] \big\vert\oper{W}\big)=\overline{\big(\rho\big\vert\Big[\big\vert \alpha)\big(\alpha\big\vert\otimes\mathcal{I}\Big] \big\vert\oper{W}\big)}.\label{eq:r-alpha-conj}
	\end{equation}
In other words, with $\big(\rho\big\vert\Big[\big\vert \alpha)\big(\alpha\big\vert\otimes\mathcal{I}\Big] \big\vert\oper{W}\big)=r_\alpha e^{i\phi_\alpha}$ we get
	\begin{equation}
	    e^{i\phi_{\alpha^\dagger}}=e^{-i\phi_\alpha}. \label{eq:phi-alpha-conj}
	\end{equation}
Consequently, we obtain
	\begin{align*}
		\Im w_\rho\cv{t} &= -\dfrac{i}{2}\cv{w_\rho\cv{t} - \overline{w_\rho\cv{t}}}\\
		&=-\dfrac{i}{2}\sum_{\alpha}\cv{e^{\lambda_\alpha t}r_\alpha e^{i\phi_\alpha} - e^{\overline{\lambda}_\alpha t}r_\alpha e^{-i\phi_\alpha}}\\
		&=-\dfrac{i}{2}\left(\sum_{\alpha}e^{\lambda_\alpha t}r_\alpha e^{i\phi_\alpha} -\sum_{\alpha} e^{\lambda_{\alpha^\dagger} t}r_{\alpha^\dagger} e^{i\phi_{\alpha^\dagger}}\right).
	\end{align*}
The second sum is equal to the first sum because its summands are just permutations of other summands. 

The properties mentioned above also lead to the classification given in Table~\ref{tab:classification}. For the case when $\cket{\alpha}$ is Hermitian, we have that $\omega_\alpha=0$, and consequently from Eq.~\eqref{eq:r-alpha-conj} the quantity $\big(\rho\big\vert\Big[\big\vert \alpha)\big(\alpha\big\vert\otimes\mathcal{I}\Big] \big\vert\oper{W}\big)$ is real. Hence, $\phi_\alpha$ can be either $0$ or $\pi$ for the Hermitian eigenelements. This scenario is relevant for classes I and II (class I is special because only there the eigenvalue is $0$).

For class III, where $\cket{\alpha}$ and its complex conjugate $\cket{\alpha^\dagger}$ are not Hermitian, one can group eigenelements of this type as $L=\cv{d^2-K-1}/2$ pairs.
For this class we will mark the element with the tilde symbol and use an index $j$ in place of $\alpha$ for the sake of clarity. 
Let $\tilde{\lambda}_j$ denote the eigenvalue of $\cket{j}$ for $j=1,\ldots,L$ if it is in the upper plane, and $\tilde{\lambda}_{2j}$ denote its complex conjugate, which is clearly an eigenvalue of $\cket{2j}.$ Also, let us express the argument $\tilde{\varphi}_j$ in the set $\cv{-\pi,\pi}\backslash\cvc{0}.$ With these conventions, the conditions in Table~\ref{tab:classification} for Class III follow from expressions Eqs.~\eqref{eq:r-alpha-conj}--\eqref{eq:phi-alpha-conj}.

\bibliographystyle{apsrev4-1}
%\bibliography{reference.bib}
\bibliography{main.bbl}

\end{document}